\documentclass[aps,pra,twocolumn]{revtex4}
\usepackage{graphicx}    
\usepackage{amsmath}
\DeclareMathOperator\erf{erf}
\DeclareMathOperator\erfc{erfc}

\begin{document}

\title{Anti-Diffusion in Continuous Opinion Dynamics}

\author{Moorad Alexanian and Dylan McNamara}

\affiliation{Department of Physics and Physical Oceanography\\
University of North Carolina Wilmington\\ Wilmington, NC
28403-5606\\}

\date{\today}

\begin{abstract}
Considerable effort using techniques developed in statistical physics has been aimed at numerical simulations of agent-based opinion models and analysis of their results. Such work has elucidated how various rules for interacting agents can give rise to steady state behaviors in the agent populations that vary between consensus and fragmentation. At the macroscopic population level, analysis has been limited due to the lack of an analytically tractable governing macro-equation for the continuous population state. We use the integro-differential equation that governs opinion dynamics for the continuous probability distribution function of agent opinions to develop a novel nonlinear partial differential equation for the evolution of opinion distributions. The highly nonlinear equation allows for the generation of a system of approximations. We consider three initial population distributions and determine their small-time behavior. Our analysis reveals how the generation of clusters results from the interplay of diffusion and anti-diffusion and how initial instabilities arise in different regions of the population distribution.
\end{abstract}

\maketitle {}

\section{Introduction}

Statistical physics forms the basis of statistical mechanics that successfully dealt with the approach to equilibrium of dilute gases as pioneered by Ludwig Boltzmann. This effort represented a seminal work for the establishment of atomism. Presently, statistical physics also forms the foundation of a large segment of theoretical investigations into social dynamics, the interactions of individuals as elementary units in social structures. The research in this area encompasses a multitude of topics from opinion and cultural and language dynamics to crowd behavior, hierarchy formation, human dynamics, and social spreading \cite{CFL09}.  The interest in many of these areas is to investigate the possibility that social modeling could become predictive.

Much of the work on the dynamics of opinion formation has focused primarily on the voter model \cite{HO75}, a framework whereby agents with discrete opinions, or votes, interact and evolve their current state opinions, often in a spatially explicit domain.  Very recent work has added attributes to the agents in the typical voter model to explore the impact of coupling the dynamics of these new attributes with the opinion dynamics  \cite{ WCMV17}.  A significant thread of the work exploring the voter model has been to see how time varying network connections between agents changes the typical consensus outcomes in the model \cite{HN06}.  Recent extensions in this direction have investigated the effect of zealots, those with fixed options, on the formation and spread of opinions within time varying social networks \cite{KWDD17}.

While the voter model has received considerable attention, many real world examples of social opinion dynamics evolve opinions that can be considered as taken from a continuous spectrum of opinion values, for example, political orientation.  The early investigations of the evolution of continuous opinions \cite{DF00,HK00} and the large majority of follow up work have relied on numerical simulations of agent interactions to explore variations between consensus, fragmentation, and polarization in steady state behavior of opinions.  A rate equation, a sort-of Boltzmann-like integro-differential equation, governing the continuously valued population of opinions can be represented analytically without appealing to individual agents.  Numerical simulation of this rate equation \cite{NKR03} reveal bifurcations of consensus into fragmented clusters as the interaction range of the population decreases.  Despite the existence of the rate equation, the analytical insight gained from the equation itself does not add to our understanding of how the continuous population of opinions evolves.  It is simply an analytic expression of agent interactions.  Here we extend the utility of the rate equation by developing a related partial differential equation for the opinion population and analytically explore the early evolution of the population from a variety of initial conditions.

This paper is arranged as follows. In Sec. II, we review exact solutions of the linear, partial differential equation that characterizes ordinary diffusion. In particular, the possible analytical continuation of the solutions for negative values of the time, that is, the anti-diffusion region. In Sec. III, we convert the Boltzmann-like integro-differential equation governing the continuous opinion dynamics into a nonlinear, partial differential equation expressed as an infinite series whereby the lowest term constitutes nonlinear anti-diffusion. The addition of further nonlinear terms constitutes a system of approximations of the original exact equation. In Sec. IV, we obtain exact power series solutions of the approximate nonlinear differential equations, where the interplay of anti-diffusion and ordinary diffusion underlies the instabilities that give rise to the nucleation and annihilation of clusters.

\section {Linear Diffusion Equation}

It is useful to consider first exact solutions of the usual one-dimensional, linear diffusion equation,
\begin{equation} \label{Diff}
\frac{\partial \psi(x,t)}{\partial t} = D \hspace{.05in} \frac{ \partial^2 \psi(x,t)}{\partial x^2},
\end{equation}
for three differing initial conditions to illustrate the analytic properties of $\psi(x,t)$. The exact solutions of the following initial conditions will allow us to study the analytical properties of $\psi(x,t)$ as a function especially of $t$, which will allow us to possibly analytically continue the solutions for $t<0$ and thus study anti-diffusion.

\subsection{Gaussian}
Consider first the Gaussian initial condition
\begin{equation}
\psi_{G}(x,0) =\alpha/\sqrt {\pi}\hspace{.06in} \exp(-\alpha^2 x^2),
\end{equation}
then the solution of (\ref{Diff}) is given by

\begin{equation} \label{Gsol}
\psi_{G}(x,t)= \frac{\alpha}{\sqrt{\pi} \sqrt{1+4\alpha^2 Dt}}\hspace{.05in} \exp{(-\frac{\alpha^2 x^2}{1+4\alpha^2 Dt} )}.
\end{equation}
Note that $\psi_{G}(x,t)$ is an entire function of $x$ but has both a branch point and an essential singularity at $t=-(4\alpha^2D)^{-1}$. Also, note that the initial condition $\psi_{G}(x,0)$ is an analytic function of $x$ and so it possesses an arbitrary number of partial derivatives. It is interesting that the function $\psi_{G}(x,t)$ can be continued analytically for negative values of the time $t$, that is, the diffusion equation can be converted into an anti-diffusion equation, which is valid for times $0>t> - (4\alpha^2D)^{-1}$. Hence, $\psi_{G}(x,t)$ starts off as a Dirac delta function at $t= -(4\alpha^2D)^{-1}$, diffuses into the Gaussian $\psi_{G}(x,0)$ at $t=0$ and continues to diffuse for $t\geq 0$ as given by (3).

\subsection{Square}

Next consider the initial condition
\begin{equation}\label{Sq_init}\psi_{S}(x,0)=\left\{\begin{array}{ll}
1/(2a) & \mbox{if $-a<x<a$}\\
0    & \mbox{otherwise}.
\end{array}
\right.
\end{equation}
The solution of the diffusion equation (\ref{Diff}) is
\begin{equation} \label{Sqsol}
 \psi_{S}(x,t)= \frac{1}{4a} \Big{(}\erf(\frac{a+x}{\sqrt{4Dt}}\Big{)} +\erf(\frac{a-x}{\sqrt{4Dt}})\Big{)},
\end{equation}
where the error function is $\erf(z)= \frac{2}{\sqrt{\pi}}\int_{0}^{z} e^{-t^2} dt$, which is an entire function of $z$. The asymptotic series of $\textup{erf} (z)$ for $z\rightarrow \infty$ shows that solution (\ref{Sqsol}) has a branch point and an essential singularity at $t=0$ and so $ \psi_{S}(x,t)$ cannot be continued analytically for negative values of $t$.

\subsection{Lorentzian}
The Lorentzian initial condition is given by
\begin{equation} \label{Lor_init}
\psi_{L}(x,0)= \frac{1}{\pi} \frac{b}{x^2+b^2}.
\end{equation}
The solution of the linear diffusion equation (\ref{Diff}) is given by
\begin{equation}
\psi_{L}(x,t)= \frac{1}{\sqrt{4\pi Dt}} \hspace{0.03in}\Re \Big{(}w\big{(}\frac{-x+ib}{\sqrt{4 Dt}}\big{)}\Big{)},
\end{equation}
where
\begin{equation}
w(z)= \frac{2iz}{\pi} \int_0^\infty \frac{e^{-t^2}}{z^2-t^2}=e^{-z^2}\erfc(-iz) \hspace{.08in}  (\Im z>0)
\end{equation}
is the Faddeeva function \cite{AS64} with series expansion
\begin{equation}
w(z)= \sum_{n=0}^{\infty} \frac{(iz)^n}{\Gamma(n/2+1 )} ,
\end{equation}
and
\begin{equation}
\Re w(z)=\frac{1}{2}\big{(}    w(z) + w(-\bar{z})   \big{)}.
\end{equation}
with $\bar{z}$ the complex conjugate of $z$. The Faddeeva function is analytic at $z=0$ and approaches unity as $z\rightarrow 0$. It is interesting that $\psi_{L}(x,t) \approx 1/\sqrt{4\pi Dt}$ as $t \rightarrow \infty$. Exactly the same asymptotic behaviors are obtained for both $\psi_{G}(x,t)$ in (\ref{Gsol}) and $\psi_{S}(x,t)$  in (\ref{Sqsol}) for large $t$. On the other hand, an expansion of $\psi_{L}(x,t)$ about $t=0$ is actually given by the asymptotic series
\begin{equation} \label{Lsol}
\psi_{L}(x,t)\sim \frac{1}{\pi} \frac{b}{x^2+b^2} + \sum_{m=1}^{\infty} A_{m} t^m,
\end{equation}
where
\begin{equation} \label{Am}
A_{m}= \frac{(2m-1)!! (2D)^m}{\pi(x^2+b^2)^{2m+1}} \sum_{l=0}^{m} \frac{(-1)^l (2m+1)!b^{2l+1} x^{2m-2l}}{(2m-2l)!(2l+1)!}.
\end{equation}
Now, for $m=1$, (\ref{Am}) gives
\begin{equation}
A_{1}=\frac{2bD}{\pi} \frac{3x^2-b^2}{(x^2+b^2)^3}
\end{equation}
and so the probability $\psi(x,t)$ decreases for $|x|<b/\sqrt{3}$ and increases for $|x|>b/\sqrt{3}$  with increasing small values of the time $t$ for normal diffusion. However, the opposite behavior is the case for anti-diffusion, which for $t\geq0$ is governed by Eq. (\ref{Lsol}) with $t$ replaced by $-t$.

\section{Continuous Compromise or Evolution of Opinions Model}

Consider the Boltzmann-like equation for the evolution of the positive distribution $P(x,t)$ \cite{NKR03,BL12,CFL09}
\[
\frac{\partial}{\partial t} P(x,t)= \int \int_{|x_{1}-x_{2}|<a} dx_{1} dx_{2} P(x_{1},t)  P(x_{2},t)
\]
\begin{equation} \label{Rate_eq}
 \times[\delta (x-\frac{x_{1}+x_{2}}{2})-\delta (x-x_{1})],
\end{equation}
where
\begin{equation}
\int_{-\infty}^{+\infty} \textup{d}x  P(x,t) =1.
\end{equation}
Using the Heaviside step function
\begin{equation}
\theta(x)=\left\{\begin{array}{ll}
1 & \mbox{ $x\geq 0$}\\
0 & \mbox{ $x<0$}
\end{array}
\right.
\end{equation}
the evolution equation (\ref{Rate_eq}) can be written
\[
\frac{\partial}{\partial t} P(x,t) = 2 \int_{x-a/2}^{x+a/2} \textup{d}x_{2}  P(2x-x_{2},t)P(x_{2},t)
\]
\begin{equation} \label{new_Rate_eq}
-P(x,t) \int_{x-a}^{x+a} \textup{d}x_{2}  P(x_{2},t).
\end{equation}
It is clear from (\ref{new_Rate_eq}) that $P(x,t)$ is actually a function of the scaled variables $x/a$ and $at$. Therefore, there is no loss of generality if we set $a=1$.

On expanding the integration variable $x_{2}$ about the point $x$ in the integrands in (\ref{new_Rate_eq}) and integrating over $x_{2}$, one obtains
\[
\frac{\partial}{\partial t} P(x,t) = 2\sum_{l=1}^{\infty} \frac{1}{2^{2l}(2l+1)}\sum_{n=0}^{2l}  \frac{(-1)^n}{n!(2l-n)!}\frac{\partial^n}{\partial x ^n} P(x,t)
\]
\begin{equation} \label{newnew_Rate_eq}
\times \frac{\partial^{2l-n}}{\partial x^{2l-n}} P(x,t) -2P(x,t)\sum_{l=1}^{\infty}\frac{1}{(2l+1)(2l)!}\frac{\partial^{2l}}{\partial x^{2l}} P(x,t).
\end{equation}

Successive approximations for $P(x,t)$ from (18) are obtained by truncating the series over the dummy variable $l$ to a particular integer value $l_{max}$. For instance, for $l_{max}=1$, one obtains from (\ref{newnew_Rate_eq}) the nonlinear partial differential equation
\[
\frac{\partial}{\partial t} P(x,t)= -\frac{1}{6}\frac{\partial}{\partial x} \Big{(} P(x,t)\frac{\partial}{\partial x} P(x,t) \Big{)}
\]
\begin{equation}
=-\frac{1}{12}\frac{\partial^2}{\partial x^2}P^2(x,t),
\end{equation}
which is a diffusion equation with the negative diffusion constant $-(1/6) P(x,t)$ and a correction term to the diffusion equation given by $-\frac{1}{6}(\frac{\partial}{\partial x} P(x,t))^2$. Note that values of $x$ where $P(x,t)$ has a maximum, $P(x,t)$ will further increase as time goes on and where $P(x,t)$ has a minimum, $P(x,t)$ will further decrease as time goes on. This is the effect of anti-diffusion.

For $l_{max}=2$, Eq. (\ref{newnew_Rate_eq}) yields
\[
\frac{\partial}{\partial t} P(x,t)= -\frac{1}{6}\frac{\partial}{\partial x} \Big{(} P(x,t)\frac{\partial}{\partial x} P(x,t)  +\frac{7}{80}P(x,t)\frac{\partial^3}{\partial x ^3} P(x,t)
\]

\begin{equation} \label{l2_eq}
-\frac{3}{80}\frac{\partial}{\partial x } P(x,t)\frac{\partial^2}{\partial x ^2} P(x,t)\Big{)},
\end{equation}
with the first and third terms in (\ref{l2_eq}) representing the anti-diffusion constant $-(1/6) \Big{(}P(x,t) -(3/80) \hspace{.05in} \partial^2 P(x,t)/\partial x ^2\Big{)}$ while the second term represents a correction to the diffusion equation itself. Note that contrary to the case with $l_{max}=1$, the case $l_{max}=2$ is more complicated and so the effect of anti-diffusion is not as clear. Similarly, one can keep three or four terms in the RHS of (18), viz., $l_{max}=3$ and $l_{max}=4$, respectively.  The numerical examples in Sec. IV for $P(x,t)$ are all based on the $l_{max}=4$ case and to second order in the time $t$ in (21) (see below), which illustrates the more complicated interplay between ordinary diffusion and anti-diffusion by the added terms to both the diffusion constant and to the generalized diffusion equation. Generalizations to the linear diffusion equation have been investigated by previous authors \cite{VVB97,BL06}.

One may obtain $P(x,t)$ as a power series expansion in the time $t$, which may be an asymptotic series,
\begin{equation} \label{series_eq}
P(x,t)=\sum_{n=0}^{\infty} \frac{t^n}{n!}\hspace{.1in}\frac{\partial^n}{\partial t ^n} P(x,t)\Big{|}_{t=0}
\end{equation}
with the aid of (\ref{newnew_Rate_eq}) for a given $l_{max}$ and initial condition $P(x,0)$, provided that $P(x,0)$ is an analytic function of $x$. If the initial condition $P(x,0)$ is an even function of $x$, then so will $P(x,t)$ be an even function of $x$.

In what follows, we will consider solutions of (18) with $l_{max}=4$ for the three initial conditions considered in Sec. II and to second order in the time variable $t$ in Eq. (\ref{series_eq}).

\section {Numerical Solutions}

The nonlinear equation (18) together with the initial condition $P(x,0)$ can be used to generate exactly the power series solution (21). It is clear that the series may be an asymptotic series and we will be obtaining the exact, first few terms of the power series in $t$. We will represent the solutions in the form
\begin{equation}
P(x,t)= P(x,0)\Big{[}1+g(x,t)\Big{]},
\end{equation}
where we can calculate exactly the first few terms in the power series in $t$ of $g(x,t)$,

\subsection{Gaussian}

\begin{figure}
\begin{center}
   \includegraphics[scale=0.6]{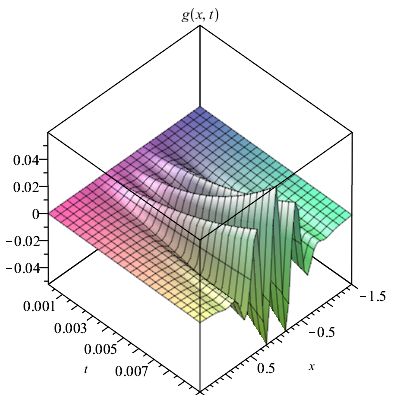}
\end{center}
\label{fig:thefig1}
  \caption{Plot of $g(x,t)$ of Eq. (22) corresponding to $l_{max}=4$ and to second order in $t$ for the Gaussian initial condition $\psi_{G}(x,0)$ with $\alpha=1.95$.  }
\end{figure}

Consider first the case of the Gaussian initial condition $\psi_{G}(x,0)$. Fig. 1 shows $g(x,t)$ for $\alpha=1.95$.

\subsection{Fermi-Dirac}

One has that the initial condition $\psi_{S}(x,0)$ in (4) cannot be used in conjunction with our system of nonlinear equations (18) in order to generate the series expansion (21) since the series does not exist, even as an asymptotic series, owing to the highly singular derivatives associates with the initial condition (4) at the Fermi surface $x=\pm x_{0}$. Instead, consider the initial condition associated with the Fermi-Dirac distribution
\begin{equation}\label{FD_init}
P(x,0)= \frac{2}{\beta}\hspace{.02in} \frac{\ln(e^{\beta x_{0}}+1)}{e^{\beta(|x|-x_{0})}+1},
\end{equation}
where the constants $x_{0}$ and $\beta$ are positives. We have obtained the exact series solution to second order in the time variable $t$ in (21). Fig. 2 shows the corresponding plot for $g(x,t)$ for the constants $\beta=50$ and $x_{0}=0.5$. Note the very small range of values of the time $t$ since one is dealing in this case with very large values for the derivatives of the initial condition. Also, $P(x,t)$ is essentially constant for $-0.4\lesssim x \lesssim 0.4$ owing to the flatness of the initial condition (23).

\begin{figure}
\begin{center}
   \includegraphics[scale=0.6]{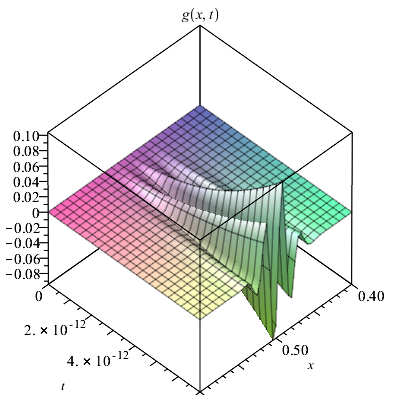}
\end{center}
\label{fig:thefig2}
  \caption{Plot of $g(x,t)$ of Eq. (22) corresponding to $l_{max}=4$ for the Fermi-Dirac initial condition (23) to second order in $t$ in the series (21) with $\beta=50$ and $x_{0}=0.5$. Note  that $g(-x,t)=g(x,t)$.   }
\end{figure}

\subsection{Lorentzian}
Next we have the Lorentzian initial condition (\ref{Lor_init}). We have calculated the exact power series solution to second order in the time variable $t$ for $l_{max}=4$. Fig. 3 shows the corresponding plot for $g(x,t)$ for the constant value $b=1$ in (6).

\begin{figure}
\begin{center}
   \includegraphics[scale=0.6]{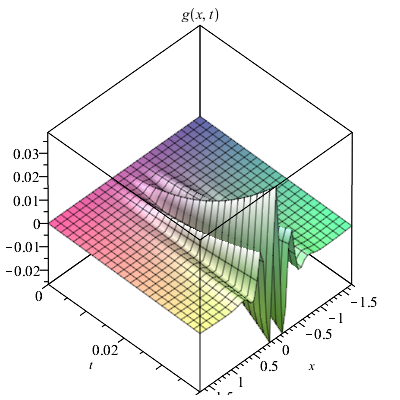}
\end{center}
\label{fig:thefig3}
  \caption{Plot of $g(x,t)$ of (22) corresponding to $l_{max}=4$ for the Lorentzian initial condition (6) with $b=1$. }
\end{figure}

In the above three cases, the temporal evolution of $P(x,t)$ results from the interplay between the processes of diffusion and anti-diffusion. Figs. 1 and 3 show that the diffusion about $x=0$ gives rise, via the process of anti-diffusion, in the formation of side band peaks that may become part of the final opinion clusters. These results, based on a few terms of the series (\ref{series_eq}), suggest that if such agents within distinct clusters exceeds the threshold of influence, then only agents within some cluster may interact and so one obtains a convergence of opinions. Accordingly, the final opinion configuration would result in a succession of Dirac $\delta$-functions. Note that in Fig. 2, the instability, owing to the diffusion and anti-diffusion processes, occurs at the Fermi surface ($x_{0}=0.5$) of the initial Fermi-Dirac distribution (\ref{FD_init}), around which the formation of peaks takes place.

\section{summary and discussion}

The reduction of the integro-differential equation governing the dynamics of opinion formation to a highly nonlinear partial differential equation allows for the generation of an approximation scheme whereby higher order derivatives may be neglected. Initial Gaussian, Fermi-Dirac, and Lorentzian distributions are solved exactly for small times with clear evidence of cluster formation owing to the instabilities generated by the interplay of diffusion and anti-diffusion.

The entirety of previous work on opinion formation focuses on steady state behaviors of a population of agents as explored through direct numerical simulation of agent interactions. In our analytic work, we are limited to exploring initial formations in the opinion distribution. The insight gained with our analysis, rather than being focused on the complete evolution of the opinion distribution, is instead aimed at the governing partial differential equations that have been developed here and the resulting initial instability. Specifically, in the first order equation (19), the time rate of change of opinions is related to a gradient in an opinion flux given by,
\begin{equation}
F_{0}=P(x,t)\frac{\partial}{\partial x} P(x,t).
\end{equation}

As our diffusion equation has a negative sign, there will be a flux of opinions toward prominent opinions, and the flux is stronger the larger the prominence of the opinion, a nonlinearity that distinguishes our system from simple time reversed linear diffusion. Said another way, if a large proportion of a population has a particular opinion, they will tend to draw nearby opinions into closer agreement, and the more folks that come to agreement, the stronger the subsequent draw toward the rapidly dominating opinion. Knowledge of this flux and the macroscopic dynamical behavior of the distribution of populations provides testable hypothesis for scientists collecting data on opinions. In particular, rather than focusing data collection on specific agent behaviors to map to agent models of opinion formation, our work suggests that data collection focus on population distributions to test how the opinion fluxes operate.

With regard to the initial instability, the region of the population distribution that shows the first sign of instability is critically dependent on the initial shape of the distribution. For the Gaussian (Fig. 1) and Lorentzian (Fig. 3) initial distributions, the instability is initiated at the center of the distribution, whereas for the Fermi-Dirac distribution (Fig. 2) the initial instability forms at the edges of the distribution. The opinion interpretation here would be that in the former case, the moderates in the crowd initiate the eventual formation of distinct clusters while in the latter case the extreme opinions start the cluster formation.

Again, thinking of data collection efforts for social scientists, these results suggest that probing at the distribution level for signs of instability could provide useful means for testing opinion dynamics models.

\begin{newpage}
\bibliography{}

\end{newpage}
\end{document}